\documentclass{article}
\usepackage{spconf,amsmath,graphicx,hyperref}

\usepackage{enumitem}
\usepackage{booktabs}
\usepackage{multirow}
\usepackage{pifont}
\usepackage{xspace}
\usepackage{listings}
\usepackage{xcolor}
\usepackage{colortbl}
\usepackage{nicematrix}
\usepackage{afterpage}
\usepackage{textcomp}

\usepackage{algorithm}
\usepackage[noend]{algpseudocode}
\usepackage{bbm}
\usepackage{comment}

\usepackage{amssymb}

\definecolor{LightCyan}{rgb}{0.88,1,1}
\newcolumntype{a}{>{\columncolor{LightCyan}}c}
\definecolor{myred}{rgb}{0.6, 0, 0}

\title{Beyond Fixed Targets: On-Policy One-Shot Speaker Personalization for Autoregressive Text-To-Speech Models}
\name{Sangkwon Park$^{1}$, Che Hyun Lee$^{1}$, Heeseung Kim$^{2\dagger}$, Sungroh Yoon$^{1,3\dagger}$\thanks{$\dagger$ Corresponding authors}}
\address{$^{1}$ECE, Seoul National University\quad$^{2}$University of Seoul
\\ $^{3}$IPAI, AIIS, ASRI, INMC, and ISRC, Seoul National University}

\begin{document}
%
\maketitle
\begin{abstract}
One-shot personalization of autoregressive (AR) text-to-speech (TTS) models remains challenging because supervised fine-tuning (SFT) on a single reference pair rapidly degrades generation quality. We identify fixed, off-policy supervision as a key cause of failure in one-shot SFT for AR TTS, leading to blank or near-empty outputs at inference. Diversifying the training data with zero-shot pseudo-pairs eliminates this failure but yields little improvement in speaker similarity and increases word error rate because the targets remain fixed and teacher-forced. We therefore propose two on-policy approaches for one-shot speaker personalization: Group Relative Policy Optimization (GRPO) with a speaker-similarity reward and On-Policy Self-Distillation (OPSD) with soft supervision from a reference-conditioned teacher. Both on-policy methods improve speaker similarity while preserving intelligibility, with different trade-offs: GRPO achieves the strongest overall performance, whereas OPSD enables substantially faster adaptation without external reward evaluation. Further analysis confirms that these gains stem from on-policy supervision rather than target diversity, establishing on-policy training as an effective framework for one-shot AR TTS personalization.
\end{abstract}

\section{Introduction}
\label{sec:intro}

Most modern autoregressive (AR) text-to-speech (TTS) systems generate speech as sequences of discrete codec tokens. Built on pretrained language models, they provide strong linguistic modeling and support zero-shot voice cloning through in-context learning (ICL) \cite{ye2025llasa,du2024cosyvoice,liao2024fish}. Given a short reference utterance and its transcript, the model uses them as the input prompt and generates arbitrary text in the same voice.

\begin{figure}[!t]
\centering
\includegraphics[width=8.7cm]{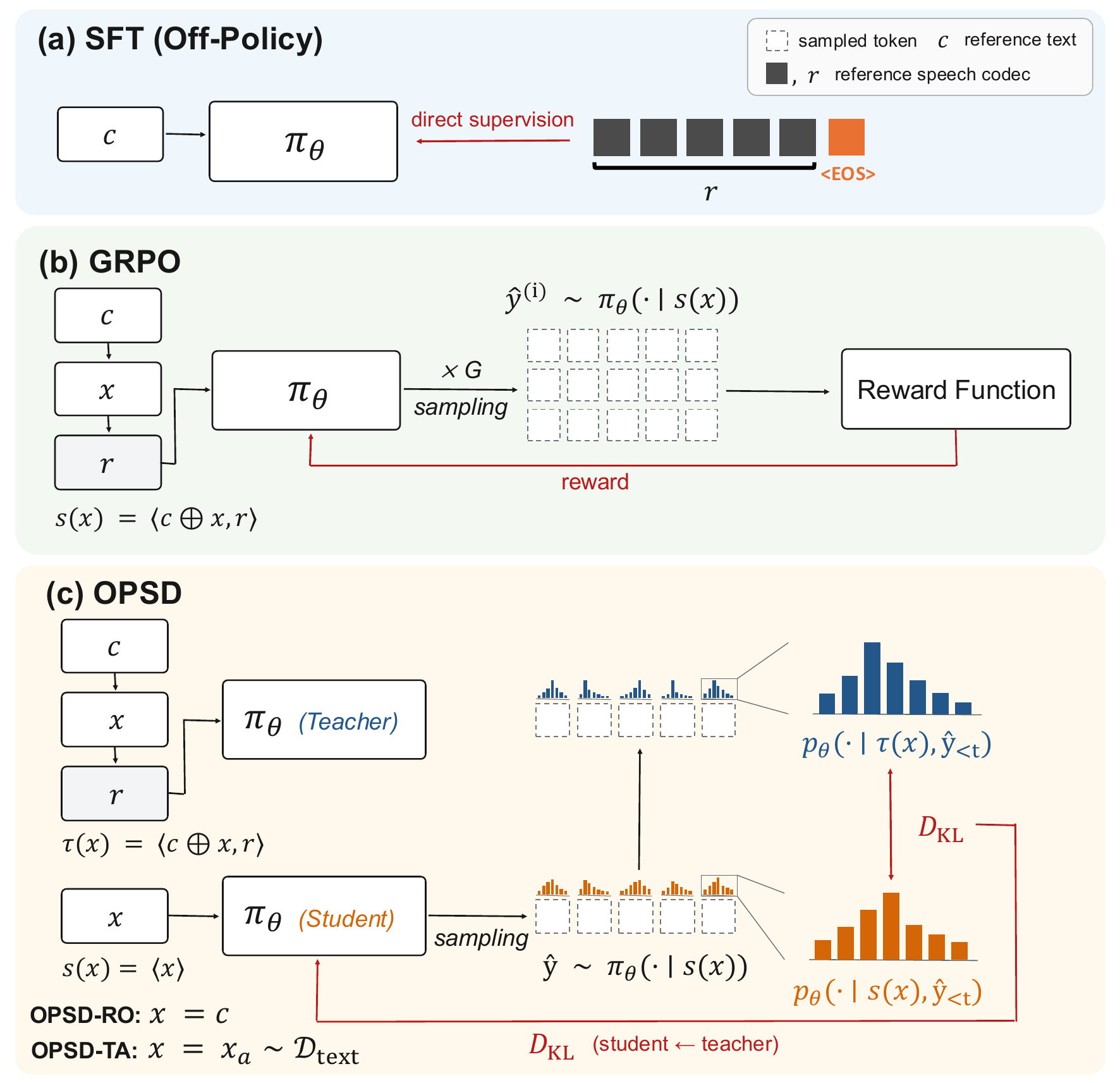}
\vskip -0.1 in
\caption{Overview of off-policy SFT and proposed on-policy methods for one-shot personalization of autoregressive TTS.}
\vskip -0.15 in
\label{fig:architecture}
\end{figure}

To further improve speaker personalization, a common approach is to fine-tune a pretrained TTS model using speech from the target speaker \cite{neuralvoicecloning}. In practice, obtaining long, clean reference speech is often difficult, and personalization may need to rely on only a single short utterance and its transcript. Supervised fine-tuning (SFT) on this reference pair is a natural baseline and has proven effective for diffusion-based TTS models \cite{unitspeech}. For AR models, however, SFT can substantially increase word error rate (WER) \cite{du2024cosyvoice2}. This limitation is particularly pronounced in the one-shot setting, where the model is repeatedly optimized using only a single fixed acoustic target.


We identify this fixed-target, off-policy formulation as the central obstacle to one-shot personalization of AR TTS models. As illustrated in Fig.~\ref{fig:architecture}, during SFT, the model is repeatedly trained to predict EOS immediately after the reference speech sequence. When the same reference speech is later included in the inference prompt, the adapted model consequently emits EOS immediately, producing blank or near-empty speech. As a result, speaker similarity deteriorates and WER increases substantially. Restricting the update with LoRA \cite{hu2021loralowrankadaptationlarge} does not address the underlying problem because it retains the same fixed target and teacher-forced objective.

To address this failure, we propose two on-policy methods for one-shot AR TTS personalization, adapting techniques widely used in textual LLMs. Both methods learn from trajectories sampled from the model’s current policy. (1) Group Relative Policy Optimization (GRPO) \cite{shao2024deepseekmath} samples multiple utterances for each text and updates the model using group-relative advantages derived from speaker-similarity rewards. (2) On-Policy Self-Distillation (OPSD) \cite{opsd} instead matches the student distribution to that of a reference-conditioned teacher along student-generated trajectories, requiring no external reward evaluation. We introduce Reference-Only and Text-Augmented OPSD variants that use either only the reference pair or additional unpaired texts during adaptation. We also introduce Zero-Shot Distilled SFT, a control that performs teacher-forced training on a fixed set of diverse pseudo-pairs generated by the zero-shot model, isolating the effect of text diversity from that of on-policy supervision.

Experiments show that diversifying SFT targets eliminates blank-speech generation but provides only a marginal similarity gain while increasing WER. Our on-policy methods instead improve speaker similarity while preserving intelligibility: GRPO achieves the strongest overall performance, and OPSD provides substantially faster adaptation without external reward evaluation. Our analysis identifies on-policy supervision, rather than target diversity, as the source of these gains and shows that it remains effective with a single transcript. Overall, we establish on-policy learning as a robust framework for one-shot AR TTS personalization.

\section{Method}
\label{sec:method}

\subsection{Problem Setup and Notation}
\label{ssec:setup}

Let $\pi_\theta$ be a pretrained AR TTS model that generates codec tokens $y=(y_1,\dots,y_T)$ as
$\pi_\theta(y\mid p)=\prod_{t=1}^{T}\pi_\theta(y_t\mid y_{<t},p)$,
where $p$ is the conditioning prefix. A codec decoder converts $y$ into a waveform. Given a target-speaker reference with transcript $c$ and codec tokens $r$, zero-shot voice cloning is performed by prefilling the model with $\langle c\oplus x,r\rangle$, allowing the continuation for target text $x$ to inherit the reference voice. Here, the comma separates the text segment $c\oplus x$ from the speech-prefix segment $r$; in the actual model input, these segments are delimited by modality-specific special tokens.

We study \emph{one-shot speaker personalization}, where $(c,r)$ is the only paired target-speaker observation for adaptation as illustrated in Figure~\ref{fig:architecture}. Arbitrary text $x$ may be drawn from an unpaired text pool $\mathcal{D}$, but no corresponding target-speaker speech is available.


\subsection{Off-Policy SFT and Data-Diversified Control}
\label{ssec:sft}

For a text input $u$ and acoustic target $v$, SFT minimizes
\begin{equation}
\mathcal{L}_{\mathrm{SFT}}(\theta;u,v)
=
-\frac{1}{|v|}
\sum_{t=1}^{|v|}
\log\pi_\theta
\bigl(v_t\mid v_{<t},u\bigr).
\label{eq:sft}
\end{equation}
Naive SFT applies Eq.~\eqref{eq:sft} to $(u,v)=(c,r)$ at every update. We evaluate both full-parameter SFT and LoRA, where optimization is restricted to low-rank adapters.

To isolate the effect of text diversity, we also consider Zero-Shot Distilled SFT, which trains on diverse pseudo-pairs generated by the zero-shot model. For each speaker, we sample texts $x\sim\mathcal{D}$ and synthesize a speech target for each text once as
$y\sim\pi_{\theta_0}(\cdot\mid\langle c\oplus x,r\rangle)$.
The resulting $(x,y)$ pairs are used in place of $(u,v)$ in Eq.~\eqref{eq:sft}. Although these pseudo-targets cover diverse texts and reflect the reference voice, they remain fixed offline targets throughout adaptation.

\subsection{GRPO with Speaker-Similarity Reward}
\label{ssec:grpo}

We adapt Group Relative Policy Optimization (GRPO) \cite{shao2024deepseekmath} to one-shot speaker personalization. GRPO updates the policy using relative rewards within a group of sampled rollouts. For each $x \sim  \mathcal{D}$, a group of $G$ rollouts, $\{y_i\}_{i=1}^{G} \sim \pi_{\theta_{\mathrm{old}}}\bigl(\cdot \mid \langle c \oplus x, r \rangle\bigr)$, is sampled from the previous policy. The objective is

\begingroup
\setlength{\multlinegap}{2pt}
\begin{multline}
\mathcal{J}(\theta)
=
\mathbb{E}_{x,\{y_i\}}
\Biggl[
\frac{1}{G}
\sum_{i=1}^{G}
\frac{1}{|y_i|}
\sum_{t=1}^{|y_i|}
\\
\Bigl(
\min\bigl(
\rho_{i,t}\hat{A}_i,\,
\mathrm{clip}(\rho_{i,t},1{-}\epsilon,1{+}\epsilon)\hat{A}_i
\bigr)
-
\beta\,
\mathrm{D_{KL}}
\bigl[\pi_\theta\,\|\,\pi_{\theta_0}\bigr]
\Bigr)
\Biggr],
\label{eq:grpo}
\end{multline}
\endgroup
where
$\rho_{i,t}=
\pi_\theta(y_{i,t}\mid y_{i,<t},\cdot)/
\pi_{\theta_{\mathrm{old}}}(y_{i,t}\mid y_{i,<t},\cdot)$
is the importance ratio, and
$\hat{A}_i=(R_i-\mathrm{mean}(\{R_j\}))/\mathrm{std}(\{R_j\})$
is the group-normalized advantage.

We use the reward as
$R_i=\mathrm{SIM}(y_i,r)$,
the cosine similarity between speaker embeddings of the decoded rollout and reference speech. This directly optimizes similarity to the reference speaker without requiring target-speaker speech for $x$.

\subsection{On-Policy Self-Distillation}
\label{ssec:opsd}

We also adapt On-Policy Self-Distillation (OPSD)~\cite{opsd}, a variant of On-Policy Distillation \cite{opd} that uses the same model with privileged context as the teacher. The student rollout $y$ is sampled from
$y\sim\pi_\theta(\cdot\mid s(x))$, where the context is the text $s(x)=\langle x\rangle$.
A frozen teacher $\pi_{\theta_0}$ evaluates the same trajectory with privileged reference context $\tau(x)$. The student minimizes
\begin{multline}
\mathcal{L}_{\mathrm{OPSD}}(\theta)
=
\frac{1}{|y|}
\sum_{t=1}^{|y|} \\[-2pt]
\mathrm{D}_{\mathrm{KL}}
\left(
\pi_\theta
\bigl(\cdot\mid y_{<t},s(x)\bigr)
\,\middle\|\,
\pi_{\theta_0}
\bigl(\cdot\mid y_{<t},\tau(x)\bigr)
\right).
\label{eq:opsd}
\end{multline}
This objective transfers the teacher's reference-conditioned behavior to the student on student-sampled trajectories. We propose two variants that differ in their teacher context $\tau(x)$ with linguistic coverage.



\textbf{Reference-Only OPSD.}
Reference-Only OPSD operates solely on the reference pair, with $c$ serving as both the reference transcript and the target text:
$\tau(c)=\langle c\oplus c,r\rangle$.
Because $(c,r)$ is a real target-speaker pair, the teacher distribution is anchored to real target-speaker speech matched to the target text. This provides reliable speaker supervision but restricts adaptation to the single linguistic and phonetic context of $c$.

\textbf{Text-Augmented OPSD.}
Text-Augmented OPSD extends adaptation to arbitrary texts $x_a\sim\mathcal{D}$ for which no paired target-speaker speech is available. For each text, the teacher uses
$\tau(x_a)=\langle c\oplus x_a,r\rangle$,
corresponding to the zero-shot voice-cloning configuration. Its supervision relies on the pretrained model's reference-conditioned distribution, which may reproduce the target voice imperfectly. However, this allows adaptation across a broader range of linguistic and phonetic contexts.
\begin{table}[t]
  \centering
  \caption{One-shot personalization results with LLaSA-1B on Seed-TTS-Eval.
  SFT-ZSD denotes Zero-Shot Distilled SFT; $t$ denotes the adaptation time measured per training step in seconds, and vs. SFT reports the ratio of $t$ relative to SFT.}
  \label{tab:main}

  \setlength{\tabcolsep}{4.5pt}
  \resizebox{\columnwidth}{!}{
  \begin{tabular}{lcccc}
    \toprule
    Method & WER (\%) $\downarrow$ & SIM $\uparrow$ & $t\downarrow$ & vs.\ SFT \\
    \midrule
    Zero-shot (ICL)    & 3.7            & 0.574  & --                       & --            \\
    \midrule
    SFT                & 18.2           & 0.536  & \phantom{00}5.8          & 1.0$\times$   \\
    SFT-LoRA           & 35.0           & 0.462  & \phantom{00}5.8          & 1.0$\times$   \\
    SFT-ZSD            & \phantom{0}8.3 & 0.586  & \phantom{0}14.8          & 2.6$\times$   \\
    \midrule
    OPSD-RO            & \phantom{0}4.9 & 0.608  & \phantom{0}35.3          & 6.1$\times$   \\
    OPSD-TA            & \phantom{0}3.7 & 0.609  & \phantom{0}\textbf{33.8} & \textbf{5.8}$\times$ \\
    GRPO               & \phantom{0}\textbf{3.0} & \textbf{0.647} & 262.2   & 45.2$\times$  \\
    \bottomrule
  \end{tabular}
  }
\end{table}

\section{Experiments}
\label{sec:experiments}

\subsection{Dataset}
\label{ssec:data}

We evaluate on the English subset of Seed-TTS-Eval~\cite{seedtts}.
For each speaker, the dataset provides one reference utterance and its transcript, along with evaluation texts and corresponding ground-truth speech. GRPO and Text-Augmented OPSD additionally use a fixed pool of English sentences drawn from LibriTTS~\cite{zen2019librittscorpusderivedlibrispeech} without target-speaker speech.

\subsection{Evaluation Protocol}
\label{ssec:eval}

All systems are evaluated using the same reference conditioned configuration as zero-shot personalization, where the model generates $y\sim\pi_{\theta}(\cdot\mid\langle c\oplus x,r\rangle)$ for evaluation text $x$ with temperature $0.8$, top-$p{=}1.0$, and a maximum length of 2048 tokens. WER is computed using \textit{whisper-large-v3}~\cite{whisper}.
Speaker similarity (SIM) is measured as the cosine similarity between ECAPA-TDNN ~\cite{ecapa-tdnn} embeddings of the generated and reference speech.

\subsection{Training Setup}
\label{ssec:training}

We use LLaSA-1B, a Llama-based AR TTS model that produces XCodec2 codec tokens. Each speaker is adapted independently only using a single text-speech pair from the user~\cite{ye2025llasa}.

\textbf{SFT} and \textbf{SFT-LoRA} repeatedly train on the single reference pair with a mini-batch size of $B=1$.  \textbf{Zero-Shot Distilled SFT} trains on self-generated pairs with $B=4$.
\textbf{Reference-Only OPSD} uses a single transcript per speaker, and samples $G=10$ rollouts that are optimized with $B=1$.  \textbf{Text-Augmented OPSD} samples $G=10$ rollouts for each of four text inputs, yielding a generation batch of 40 rollouts that is optimized in mini-batches of size $B=4$. \textbf{GRPO} samples $G=8$ rollouts for each of four text inputs, yielding a generation batch of 32 rollouts that is likewise optimized in mini-batches of size $B=4$. For OPSD, the token-level KL divergence is computed over the teacher's top-$K=64$ logits. All on-policy rollouts are sampled with temperature $0.9$, top-$p{=}0.95$, top-$k{=}50$, and a maximum length of 500 tokens.

\begin{figure}[t]
\centering
\centerline{\includegraphics[width=8cm]{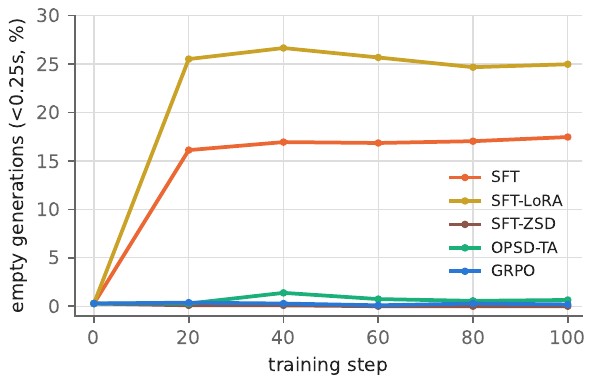}}
\caption{Ratio of empty generations per training step.}
\label{fig:empty_rate}
\vskip -0.15 in
\end{figure}

\section{Results}
\label{sec:results}

\subsection{Limitations of Off-Policy SFT}
\label{ssec}

As shown in Table~\ref{tab:main}, the conventional SFT formulation is unsuitable for one-shot personalization: WER increases substantially, and SIM falls below the zero-shot baseline. The model frequently generates blank or near-empty speech, as shown in Fig.~\ref{fig:empty_rate}. This behavior follows directly from its objective. Each update maximizes the likelihood of the reference tokens $r$ followed by EOS given the reference transcript $c$. At inference, the prompt $\langle c\oplus x,r\rangle$ ends with the same $r$, causing the model to emit EOS immediately after the reference prefill. Removing the reference prompt is not a viable workaround, as it substantially worsens both WER and SIM.

Attempts to mitigate overfitting with LoRA do not address this objective-level problem and instead lead to even greater performance degradation and a higher empty-generation rate. Zero-Shot Distilled SFT replaces the single ground-truth target with diverse but fixed pseudo-targets generated by the zero-shot model. This diversification eliminates the termination failure, but yields only a marginal improvement in speaker similarity while substantially increasing WER. 
These results motivate our shift from data diversification within SFT to on-policy supervision.

\subsection{On-Policy Personalization}
\label{ssec:onpolicy}

Our GRPO and OPSD methods directly address the limitation above by learning from current-policy rollouts. Both avoid fitting fixed hard targets: GRPO uses reward-weighted policy updates, while OPSD matches soft next-token distributions from a reference-conditioned teacher. As shown in Table~\ref{tab:main}, this on-policy formulation preserves intelligibility while substantially improving speaker similarity. Both OPSD variants improve SIM with only minor changes in WER, whereas GRPO achieves the highest SIM and lowest WER. 
Table~\ref{tab:mos} further shows that our methods achieve naturalness comparable to zero-shot synthesis without sacrificing speech quality.
The methods differ substantially in computational cost. OPSD adapts approximately $8\times$ faster than GRPO while achieving comparable speaker similarity, as it avoids repeated vocoding and external reward evaluation. These methods thus offer different advantages: GRPO prioritizes peak performance, whereas OPSD enables faster, reward-free adaptation.

\begin{table}[t]
  \centering
  \caption{5-scale NMOS (Naturalness Mean Opinion Score) with 95\% CI, based on 120 evaluations from 93 evaluators.}
  \label{tab:mos}
  \setlength{\tabcolsep}{2.5pt}
  \resizebox{\columnwidth}{!}{%
  \begin{tabular}{@{}lccccc@{}}
    \toprule
    & Zero-Shot & SFT & OPSD-TA & GRPO & Ground Truth \\
    \midrule
    NMOS $\uparrow$
      & $3.83\pm0.16$
      & $3.42\pm0.21$
      & $3.80\pm0.15$
      & $3.87\pm0.15$
      & $4.38\pm0.09$ \\
    \bottomrule
  \end{tabular}%
  }
  \vspace{-0.1in}
\end{table}

\section{Analysis}
\label{sec:analysis}

\subsection{Why On-Policy Supervision Matters}
\label{ssec:diversity}

The failure of Zero-Shot Distilled SFT shows that data diversity alone is insufficient for effective one-shot personalization. Zero-Shot Distilled SFT, GRPO, and Text-Augmented OPSD all operate on speech drawn from the model's zero-shot cloning behavior. However, Zero-Shot Distilled SFT uses fixed samples and repeatedly treats them as hard targets. As the model changes, training on these fixed trajectories becomes increasingly off-policy, encouraging overfitting and degrading intelligibility without introducing speaker information beyond the initial zero-shot outputs.

The proposed on-policy methods instead apply supervision to trajectories sampled from the current model. This design prevents overfitting to fixed acoustic sequences and avoids memorizing the transition from the reference speech to EOS. The success of Reference-Only OPSD further validates the formulation: even with a single transcript, using the reference speech as privileged teacher context rather than as a fixed output target prevents collapse.

Notably, on-policy methods improve beyond their initial zero-shot rollouts through different supervision signals. GRPO scores current rollouts with an external speaker-similarity reward and increases the likelihood of higher-reward samples. Text-Augmented OPSD instead obtains soft reference-conditioned distributions from the teacher at states visited by the current student. Although this teacher uses the same zero-shot configuration, its distribution provides richer information than a single sampled output. Our methods can therefore refine zero-shot behavior rather than merely imitate it. This result confirms the central design choice: effective one-shot personalization requires current-policy trajectories paired with reward-based or distributional supervision, not simply a larger collection of fixed targets.



\subsection{Rollout Reuse in OPSD}
\label{ssec:sampling}


\begin{table}[t]
  \centering
  \caption{Effect of rollout reuse in OPSD on Seed-TTS-Eval.
Each rollout batch is reused for $k$ updates.}
  \label{tab:ablation_k}

  \setlength{\tabcolsep}{4.5pt}
  \resizebox{\columnwidth}{!}{
  \begin{tabular}{lcccc}
    \toprule
    Method & WER (\%) $\downarrow$ & SIM $\uparrow$ & $t\downarrow$ & vs.\ SFT \\
    \midrule
    OPSD-RO ($k{=}1$)  & \phantom{0}5.4          & 0.607          & 224.0                    & 38.6$\times$ \\
    OPSD-RO ($k{=}10$) & \phantom{0}\textbf{4.9} & \textbf{0.608} & \phantom{0}\textbf{35.3} & \textbf{6.1}$\times$ \\
    \midrule
    OPSD-TA ($k{=}1$)  & \phantom{0}\textbf{3.5} & \textbf{0.613} & 174.1                    & 30.0$\times$ \\
    OPSD-TA ($k{=}10$) & \phantom{0}3.7          & 0.609          & \phantom{0}\textbf{33.8} & \textbf{5.8}$\times$ \\
    \bottomrule
  \end{tabular}
  }
\vskip -0.15 in
\end{table}


OPSD is trained on trajectories sampled from the current policy, and each batch of trajectories can be reused for multiple parameter updates. We therefore reuse each rollout batch for $k$ consecutive updates before sampling a new batch. This introduces moderate rollout staleness but substantially reduces the cost of autoregressive generation.

As shown in Table~\ref{tab:ablation_k}, increasing the refresh interval preserves the performance of both OPSD variants while substantially reducing adaptation time. The results suggest that, in the short one-shot adaptation regime, moderate policy drift does not materially undermine OPSD supervision. Rollout reuse therefore makes OPSD substantially more efficient without requiring fresh on-policy samples at every update.



\section{Conclusion}

We identified a key limitation of fixed-target SFT for one-shot AR TTS personalization and proposed two on-policy alternatives, GRPO and OPSD. Using speaker-similarity rewards and soft teacher distributions, respectively, both learn from current-policy rollouts and improve speaker similarity while preserving intelligibility and naturalness. GRPO achieves the strongest overall performance, whereas Text-Augmented OPSD offers an efficient, reward-free alternative. Our results and analyses establish on-policy adaptation as an effective framework for one-shot AR TTS personalization.

\section{Acknowledgments}
This work was supported by the Institute of Information \& Communications Technology Planning \& Evaluation (IITP) grants funded by the Korea government (MSIT) [NO.RS-2021-II211343, Artificial Intelligence Graduate
School Program (Seoul National University); No.2022-0-00959, RS-2022-II220959], by the National Research Foundation of Korea (NRF) grant [No.2022R1A3B1077720, 2022R1A5A7083908], BK21 FOUR Program of the Education and Research Program for Future ICT Pioneers, Seoul National University in 2026.
Generative AI tools (Claude and ChatGPT) were used for coding assistance, proofreading and revision of author-written drafts throughout the paper, and figure preparation.
\section{Compliance with Ethical Standards}
The listening test was conducted through an online survey platform (Amazon Mechanical Turk) that obtained participant consent and collected only non-sensitive evaluation data.

{\fontsize{9}{9}\selectfont
\bibliographystyle{IEEEbib}
\bibliography{strings,refs}
}

\end{document}